\documentclass[graybox]{svmult}
\usepackage[utf8]{inputenc}
\usepackage[english]{babel}
\usepackage[inline]{enumitem}
\usepackage{mathtools}
\usepackage{subcaption}
\usepackage{mathrsfs}
\usepackage{hyperref}
\usepackage{color}

\usepackage{amsmath,amssymb}

\DeclareMathOperator*{\argmin}{arg\,min}

\usepackage{bm}
\usepackage[]{graphicx}
\usepackage[export]{adjustbox}
\usepackage{tikz}

\tikzset{vertex/.style={draw, circle, fill=white}}
 
\newcommand{\norm}[1]{\left\|#1\right\|}

\newcommand{\bB}{\mathbf{B}}
\newcommand{\bD}{\mathbf{D}}
\newcommand{\bA}{\mathbf{A}}
\newcommand{\bc}{\mathbf{c}}

\newcommand{\bL}{\mathbf{L}}
\newcommand{\bU}{\mathbf{U}}

\newcommand{\bu}{\mathbf{u}}
\newcommand{\bv}{\mathbf{v}}
\newcommand{\bone}{\textbf{1}}

\newcommand{\bG}{\mathbf{G}}
\newcommand{\bQ}{\mathbf{Q}}

\newcommand{\bS}{\mathbf{S}}
\newcommand{\im}{\operatorname{im}}
\newcommand{\bH}{\mathbf{H}}

\newcommand{\bef}{\mathbf{f}}
\newcommand{\bfh}{\mathbf{\hat{f}}}

\newcommand{\bI}{\mathbf{I}}

\newcommand{\bx}{\mathbf{x}}
\newcommand{\by}{\mathbf{y}}
\newcommand{\bs}{\mathbf{s}}
\newcommand{\bW}{\mathbf{W}}
\newcommand{\bY}{\mathbf{Y}}

\newcommand{\bLambda}{\boldsymbol{\Lambda}}

\newcommand{\bom}{\boldsymbol{\omega}}

\title*{Signal processing on simplicial complexes}
\author{Michael T. Schaub,  Jean-Baptiste Seby, Florian Frantzen, T. Mitchell Roddenberry, Yu Zhu, Santiago Segarra}
\authorrunning{Schaub et al.}
\institute{%
Michael T. Schaub  \at RWTH Aachen University, Germany, \email{schaub@cs.rwth-aachen.de}\newline
Jean-Baptiste Seby \at Universit\'{e} Paris-Sud, France,
\email{jbseby@mit.edu}\newline
Florian Frantzen  \at RWTH Aachen University, Germany, \email{florian.frantzen@cs.rwth-aachen.de}\newline
T. Mitchell Roddenberry \at Rice University, USA, \email{mitch@rice.edu}\newline
Yu Zhu \at Rice University, USA, \email{yz126@rice.edu}\newline
Santiago Segarra \at Rice University, USA, \email{segarra@rice.edu}}
\begin{document}

\maketitle
\vspace{-2cm}
\abstract{Higher-order networks have so far been considered primarily in the context of studying the structure of complex systems, i.e., the higher-order or multi-way relations connecting the constituent entities.
More recently, a number of studies have considered dynamical processes that explicitly account for such higher-order dependencies, e.g., in the context of epidemic spreading processes or opinion formation.
In this chapter, we focus on a closely related, but distinct third perspective: how can we use higher-order relationships to process signals and data supported on higher-order network structures.
In particular, we survey how ideas from signal processing of data supported on regular domains, such as time series or images, can be extended to graphs and simplicial complexes.
We discuss Fourier analysis, signal denoising, signal interpolation, and nonlinear processing through neural networks based on simplicial complexes.
Key to our developments is the Hodge Laplacian matrix, a multi-relational operator that leverages the special structure of simplicial complexes and generalizes desirable properties of the Laplacian matrix in graph signal processing.\\[0.3cm]}

{\small\noindent This chapter is built on the exposition of~\cite{SPTutorial}, in particular with respect to the example applications considered. Our discussion here is however more geared towards a reader with a network science background who is less familiar with signal processing. In particular, we provide additional discussion on the relations between filtering and linear dynamical processes on networks.\\[0.2cm]}

\section{Introduction}
Graphs provide a powerful abstraction for a wide variety of complex systems.
Underpinning this abstraction is the central idea of decomposing a system into its fundamental entities and their relations.
Those entities are then modelled as nodes and pairwise interactions between those entities are encoded by edges in a graph.
This highly flexible, yet conceptually simple framework has been employed in many fields~\cite{Newman2018,Easley2010}, with applications ranging from biology to social sciences.

However, the role that graphs play in such analysis can be substantially different, depending on the type of data or process under investigation~\cite{bick2021higher}. 
At the risk of oversimplification, we distinguish between two perspectives: In one case, the data are the connectivity patterns of the network itself, in the other, we aim to comprehend high-dimensional signals supported on a network.
This first viewpoint is central when modeling relational data, i.e., data corresponding to measured edges (connections, interactions) in a network, where we aim to learn about a system by finding patterns in these stochastic connections, e.g., via community detection, centrality analysis or a range of other tools.
In contrast, when studying dynamical processes on networks, we often consider the network as an arbitrary but (essentially) fixed entity, and our goal is to leverage the graph structure to understand the dynamics associated with the nodes.
This latter perspective, in which we aim to understand data \emph{supported on a graph} is the point of view we will adopt in the following.

The problem of having to understand processes and data supported on the nodes of a graph does not only arise in the context of network dynamical systems.
Another area focusing on data of this type is graph signal processing (GSP)~\cite{Sandryhaila2013,Shuman2013,Ortega2018}, which is concerned with the analysis of general signals supported on the nodes of a network.
In fact, while such signals may come in the form of a time series, e.g., from dynamical measurements at the nodes, many node signals consist of other types of attribute data supported on nodes --- sometimes referred to as node-covariates, node meta-data, or node feature data.
The goal of GSP is to extend concepts from signal processing such as the Fourier transform and the large set of filtering operations developed in signal processing to data supported on the nodes of a graph~\cite{Sandryhaila2013,Shuman2013,Ortega2018}. 
Drawing on the rich tradition of signal processing, GSP provides a range of methods to analyze and process graph supported data and, thus, has seen a steady increase in interest in the last decade.

Even though graph-based descriptions of systems have been the dominant paradigm to analyze heterogeneously structured systems and data so far (potentially augmented in terms of multiple layers or temporal dimensions), the utility of graphs for modeling certain aspects of complex systems has been scrutinized recently.
Specifically, graphs do not encode interactions between more than two nodes, even though such multi-way interactions are widespread in complex systems~\cite{Torres2020,Battiston2020,SPTutorial,bick2021higher}: assemblies of neurons fire in unison~\cite{giusti2016}, biochemical reactions typically include more than two chemical species~\cite{klamt2009}, and group interactions are widespread in social context~\cite{kee2013}. 
To represent such polyadic interactions, a number of modeling frameworks have been proposed in the literature to model such higher-order relations, including simplicial complexes~\cite{Hatcher2002}, hypergraphs~\cite{Berge1989}, and others~\cite{Frankl1995}.
Using these frameworks to analyze the organizational principles of the (higher-order edge) structure of polyadic relational data has garnered much attention in the literature lately.

In comparison to this line of work of representing and analyzing the \emph{structure} of complex multi-relational systems, the literature on dynamical processes on such higher-order network structures is still relatively sparse. 
However, there is a fast-growing body of work considering epidemic spreading, diffusion and opinion formation, among other processes, on higher-order networks -- see the recent review~\cite{Battiston2020} and references therein.
Similarly, the literature on \emph{signal processing} on higher-order networks is still nascent and received so far comparably little attention~\cite{Barbarossa2020,SPTutorial,robinson2014topological}.
Our goal in this manuscript is to present this area of signal processing on higher-order network in an accessible and coherent manner, focusing on simplicial complexes as modeling framework for higher-order network interactions.
Along the way we point out connections to the study of certain dynamical processes on (higher-order) networks and highlight avenues for future research.

The remainder of this document is structured as follows.
We assume most of our readers will be accustomed to graphs and networks, but perhaps less familiar with some of the ideas of signal processing.
Hence, we review in Section~\ref{section:SP_and_LDS} some central tenets from discrete signal processing that are relevant for our purposes, and highlight the close connection between discrete signal processing and linear dynamical systems. 
In Section~\ref{section:SP_graphs}, we use the interpretation of signal processing in terms of dynamical systems to explain how one can naturally generalize from the processing of time series to signals supported on graphs. 
In this context we introduce three example applications:
\begin{enumerate*}[label=\roman*)]
    \item signal smoothing and denoising for graph signals,
    \item signal interpolation on graphs, and
    \item nonlinear signal processing via graph neural networks.
\end{enumerate*}
In Section~\ref{section:SP_SC}, we then extend the ideas from graph signal processing to signals supported on higher-order networks, and outline central ideas for signal processing on simplicial complexes (SC).
We revisit our example applications and show how the Hodge Laplacian, a generalization of the graph Laplacian for SCs, plays a central role in the processing of signals supported on SCs.
We finally point out some further connections to the study of certain dynamical systems on graphs and simplicial complexes and then conclude with a brief discussion on future research avenues and directions.

\section{From signal processing to dynamical systems \& graphs}\label{section:SP_and_LDS}
Before embarking on our study of signal processing on simplicial complexes, we revisit key concepts from signal processing that will be instrumental for our explanations in subsequent sections.
In particular, our exposition highlights how linear signal processing and linear dynamical processes may be seen as two sides of the same coin.
% TODO
%Accordingly, many of the ideas presented in the context of signal processing on graphs may be interpreted through the lens of dynamical processes on the network, and indeed there are close connections between the diffusion processes and certain linear filtering operations.

\subsection{Linear signal processing in a nutshell}
Discrete signal processing (DSP)~\cite{oppenheim2009discrete} is concerned with the extraction of information from observed data.
Arguably one of the most elementary scenarios is that we observe some $n$ dimensional data vector $\by=\bs + \bm \eta \in\mathbb{R}^n$, which is simply an addition of some signal $\bs$ and a distortion term $\bm \eta$.
Simply put, the data consists of signal plus noise, and our goal is to extract the signal component from the observed data.

To make this task well-defined, we have to characterize more precisely what properties the signal and the noise have, i.e., we have to provide some modelling assumptions that specify characteristics of the signal versus the noise.
A typical assumption here is that the signal $\bs$ is concentrated in a particular linear subspace $\mathcal{S}\subset \mathbb{R}^n$, whereas the noise $\bm \eta$ is not localized in $\mathbb{R}^n$.
For instance, in the context of time series, we typically assume that the signal $\bs$ is varying smoothly, i.e., can be well-approximated by a linear combination of smooth basis functions.
By finding an expressive set of basis vectors such that the subspace $\mathcal S$ of ``interesting signals'' can be spanned via a (sparse) subset of basis vectors, we can filter out the noise by projecting the observed signal $\by$ onto $\mathcal S$.
The classical choice for such a basis is the collection of discrete Fourier modes~\cite{oppenheim2009discrete}.
Through their inherent characterization in terms of frequencies, Fourier modes enable us to distinguish between slowly varying, smooth signals and more rapidly oscillating signals.

The discrete Fourier transform of a signal $\by$ is defined as $\tilde \by = \mathbf{F}\by$, where $\mathbf{F}$ is the discrete Fourier transform matrix:
\begin{align}
    \mathbf{F} = \frac{1}{\sqrt{n}}\begin{bmatrix}
1 & 1 & 1 & 1 & \cdots & 1 \\
1 & \omega & \omega^{2} & \omega^{3} & \cdots & \omega^{n-1} \\
1 & \omega^{2} & \omega^{4} & \omega^{6} & \cdots & \omega^{2(n-1)} \\
1 & \omega^{3} & \omega^{6} & \omega^{9} & \cdots & \omega^{3(n-1)} \\
\vdots & \vdots & \vdots & \vdots & \ddots & \vdots \\
1 & \omega^{n-1} & \omega^{2(n-1)} & \omega^{3(n-1)} & \cdots & \omega^{(n-1)(n-1)}
\end{bmatrix} \in \mathbb{C}^{n\times n}.
\end{align}
Here $\omega = \exp({{-2i\pi}/{n}})$ is the $n$th primitive root of unity, and $i$ denotes the imaginary unit with $i^2 = -1$.
Since $\mathbf{F}$ is a unitary matrix ($\mathbf{FF}^*=\bI$), we see that the original signal can be synthesized via $\by = \mathbf{F}^*\tilde\by$, where $\mathbf{F}^*$ is the conjugate transpose of $\mathbf{F}$.
Hence, the vector of Fourier coefficients $\tilde\by$ is simply the representation of the signal $\by$ in the new Fourier basis given by the columns of $\mathbf{F}^*$, which are simply the complex conjugates of the rows of $\mathbf{F}$.
Note that the Fourier basis functions are ordered such that the first Fourier mode, the constant vector, is associated with the smallest frequency possible (frequency $0$).

Based on the Fourier transform, we can now define the concept of a linear time-invariant filter as follows\footnote{More precisely, as often encountered in DSP, we are dealing with a cyclic time-shift invariant filter.}.
Let us consider a signal vector $\bs_\text{in}$ defined via a scalar time series $s_\text{in}(t)$ at time-steps $t = 0,1,\ldots, n-1$, i.e., $\bs_\text{in} = [s_\text{in}(0), \ldots, s_\text{in}(t),\ldots, s_\text{in}(n-1)]$.
A linear time-invariant filter now consists in a transformation of the input signal $\bs_\text{in}$ via a linear operator $\bH$ to an output signal $\bs_\text{out}$:
\begin{align}\label{linear_filering}
    \bs_\text{out} = \bH \bs_\text{in},
\end{align}
such that the filter $\bH = \mathbf{F}^*\bLambda_\bH \mathbf{F}$ can be diagonalized by discrete Fourier modes, i.e., the discrete Fourier modes are eigenvectors of $\bH$.
Here $\bLambda_\bH$ is the diagonal matrix of eigenvalues of the filter $\bH$, which is the so called frequency response of the filter.
Accordingly, we can interpret the action of filter $\bH$ in~\eqref{linear_filering} in terms of three consecutive operations. 
We first express the initial signal $\bs_\text{in}$ in the Fourier basis by applying the Fourier transformation $\mathbf{F}$. 
We then modulate (amplify or attenuate) the coefficients of the signal representation in this new basis representation in a desired way by multiplying with $\bLambda_\bH$. 
Finally, we project back the output signal onto the initial basis by applying the inverse Fourier transformation $\mathbf{F}^*$.  

Via direct calculation, it can be shown that the matrix representation of any such time-invariant filter $\bH$ is a circulant matrix of the form
\begin{align}\label{eq:filtermatrix}
\bH = \begin{bmatrix}
    c_0 & c_{n-1} & c_{n-2} & \cdots & c_{1}\\
    c_{1} & c_0 & c_{n-1} & \cdots& c_{2}\\
c_{2} & c_{1} & c_0  & & c_{3}\\
\vdots & & \ddots &\ddots &\vdots\\
c_{n-1} & c_{n-2} & \cdots  & c_{1} & c_0\\
\end{bmatrix}.
\end{align}
Note that the vector $\bm \lambda(\bH)$ of eigenvalues of $\bH$, i.e., the frequency response of the filter can be calculated as:
\begin{align}
    \bm \lambda(\bH) = \sqrt{n}\cdot \mathbf{F}\mathbf{c} \quad \text{with} \quad \mathbf{c} = [c_0,\ldots,c_{n-1}]^\top,
\end{align}
which means that the eigenvalues are simply a Fourier transform of the coefficient vector $\mathbf{c}$ that defines the circulant matrix $\bH$.

The above description of signal processing in terms of a change of basis of the original (time) signal to a frequency domain representation highlights that the choice of signal \emph{representation} in terms of a basis can be crucial for processing signals --- this is a scheme we will see reoccur in the context of graphs and simplicial complexes.
However, based on this representation, the close connections between such linear filtering operations and linear dynamical systems are less apparent.
In the following subsection we thus focus on an equivalent interpretation of such filtering processes in terms of linear dynamical processes defined on (certain) graphs.

\subsection{Signal processing via linear dynamical systems on graphs}\label{section:LDS}
We now concentrate on a formulation of the above signal processing procedures in terms of linear dynamical systems in discrete time.
To this end, let us define a signal $c(t)$ based on the vector $\bc =[c(0), \ldots, c(t),\ldots, c(n-1)]^\top$, analogously as we defined $\bs_\text{in}$.
Moreover, consider the periodic extensions of both of these signals defined as $s_\text{in}^\circ(t) = s_\text{in}(t-\ell n)$ and $c^\circ(t)= c(t-\ell n)$ for $\ell\in \mathbb{Z}$.

From \eqref{eq:filtermatrix}, we observe that the filtering operation \eqref{linear_filering} may equivalently be written in terms of the linear convolution%
\footnote{Equivalently, this may also be interpreted as the cyclic convolution of the vectors $\bc$ and $\bs_\text{in}$.}
of the (periodically extended) impulse response $c(t)$ and the input signal $s_\text{in}(t)$
\begin{align}
    s_\text{out}(t) = (c * s_\text{in})(t) = \sum_{i=-\infty}^{\infty} c(i) s_\text{in}^\circ(t-i) = \sum_{i=-\infty}^{\infty} c^\circ(t-i) s_\text{in}(i).
\end{align}
Clearly, the above formula defines a linear dynamical system in which $c(t)$ plays the role of an impulse response.
The system is however not \emph{memoryless} as inputs $s_\text{in}(t')$ at times $t'\neq t$ are important for the output $s_\text{out}(t)$ of the system at time $t$.
To implement (or realize) the system we thus introduce a state vector $\bx\in\mathbb{R}^n$ which keeps track of the inputs to the system $s_\text{in}(t)$ at previous times.
Accordingly, we may express the filtering operation \eqref{linear_filering} as:
\begin{subequations}\label{eq:filter_state_space}
\begin{align}
    s_\text{out}(t) &= \bc^\top \bx(t)\\
    \bx(t+1) &= \bS\bx(t) \qquad \text{with} \quad \bx(0) = \bs_\text{in},
\end{align}
\end{subequations}
where the state transition matrix $\bS$ takes the special form of a (circular) shift 
\begin{align}
    \bS = \begin{bmatrix}
            0 & 0 & 0 & \ldots & 0 & 1 \\
            1 & 0 & 0 & \ldots & 0 & 0 \\
            0 & 1 & 0 & & 0 & 0 \\
            \vdots & \vdots & \ddots & \ddots & \ddots & 0 \\
            0 & 0 & \ldots & 1 & 0 & 0 \\
            0 & 0 & \ldots & 0 & 1 & 0
        \end{bmatrix}.
\end{align}

Note that the above realization of the system may not be minimal and we may be able to implement our system with fewer states.
However, the key idea of representing a linear system in terms of a set of internal states which are coupled by a shift operator is the aspect that is central to our further developments.
In particular, using the shift operator we can compactly summarize the above filtering operation in vector form:
\begin{align}\label{eq:shift_convolution}
    \bs_\text{out} = \sum_{k=0}^{n-1} c_k \bS^k \bs_\text{in} = \bH \bs_\text{in},
\end{align}
which provides us with yet another way to express the output of our dynamical system.
Note that this implies, in particular, that the filter $\bH$ can be constructed from linear combinations of the simpler shift operator $\bS$.
Further, the Fourier basis $\mathbf{F}^*$ diagonalizes also the shift operator $\bS$.

\begin{figure}[tb!]
    \centering

    \begin{tikzpicture}
        \begin{scope}
            \node[vertex, label=270:$x_1$] (0) at (0, 2) {};
            \node[vertex, label=270:$x_2$] (1) at (2, 2) {};
            \node[vertex, label=270:$x_3$] (2) at (4, 2) {};
            \node[vertex, label=270:$x_4$] (3) at (6, 2) {};
            \node (4) at (7.5, 2) {$\cdots$};
            \node[vertex, label=270:$x_n$] (5) at (9, 2) {};

            \draw[->] (0) -- (1);
            \draw[->] (1) -- (2);
            \draw[->] (2) -- (3);
            \draw[->] (3) -- (4);
            \draw[->] (4) -- (5);
            \draw[->] (5) to [out=160, in=20] (0);
        \end{scope}

        %\begin{scope}
            %\node[vertex, label=270:$x_n$] (0) at (0, 0) {};
            %\node[vertex, label=270:$x_1$] (1) at (2, 0) {};
            %\node[vertex, label=270:$x_2$] (2) at (4, 0) {};
            %\node[vertex, label=270:$x_3$] (3) at (6, 0) {};
            %\node (4) at (7.5, 0) {$\cdots$};
            %\node[vertex, label={270:$x_{n-1}$}] (5) at (9, 0) {};

            %\draw[->] (0) -- (1);
            %\draw[->] (1) -- (2);
            %\draw[->] (2) -- (3);
            %\draw[->] (3) -- (4);
            %\draw[->] (4) -- (5);
            %\draw[->] (5) to [out=160, in=20] (0);
        %\end{scope}
    \end{tikzpicture}

    \caption{\textbf{Interpretation of the cyclic shift operator as a linear dynamical system on a graph.} 
    The cyclic shift operator $\bS$ may be interpreted as a linear diffusion $\bx(t+1)=\bS\bx(t)$ on a directed cycle graph of size~$n$.}
    \label{fig:cycle-graph}
\end{figure}
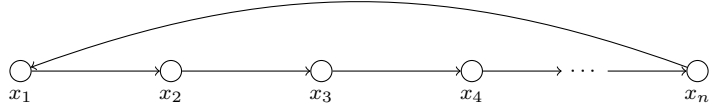

The rewriting \eqref{eq:shift_convolution} suggests an interpretation of the filtering operation as a linear process on a graph as illustrated in Figure~\ref{fig:cycle-graph}.
Specifically, we may interpret the shift operator $\bS$ as a cyclic graph $\mathcal{G}_C$ and associate each node with a state (time) of our dynamical system (see Figure~\ref{fig:cycle-graph}). 
Each iteration of the dynamical system $\bx(t+1) = \bS \bx(t)$ can be interpreted as a linear filtering operation.
Since a combination of linear filters is linear, the output $\by = \sum_t c_t \bx(t)$ is linear and can thus also be interpreted in terms of a dynamical system on a cyclic graph.
To summarize, a linear time-shift invariant filter can be interpreted as a weighted sum of the states of a linear dynamical system on a cyclic graph. 

\section{Signal processing on graphs}\label{section:SP_graphs}
In this section, we provide a short introduction to graph signal processing (GSP), which extends the ideas of signal processing for time series or images to the processing of signals supported on general graphs.
A key insight underpinning GSP is that the filtering operation~\eqref{eq:shift_convolution} can be generalized from a cyclic graph to any graph, by defining an appropriate shift operator compatible with the structure of the graph on which the signals are supported.

\subsection{Graph signals, Fourier transforms, and filters}
To introduce the ideas of graph signal processing mathematically, we will first recall some preliminary definitions and notation for graphs and signals supported on graphs.
For simplicity, we will concentrate on undirected graphs in the ensuing discussion, as this setup can be directly generalized to simplicial complexes. 
However, graph signal processing may also be considered in the context of directed graphs~\cite{furutani2019graph,marques2020signal}.

We define an undirected graph $\mathcal{G}$ by a set of nodes $\mathcal{V} = \{v_1, \cdots, v_n\}$ with cardinality $|\mathcal{V} | = n$ and a set of edges $\mathcal{E}$, i.e., a collection of unordered pairs of nodes, with cardinality $|\mathcal{E}| = m$ . 
A graph signal $s : \mathcal{V} \rightarrow \mathbb{R}$ is a mapping that assigns to each node $i\in\mathcal{V}$ a real-valued scalar.
Such a graph signal may thus be suitably represented by a vector $\bs\in \mathbb{R}^n$.

For computational purposes, we encode the structure of the graph $\mathcal G$ by an adjacency matrix $\bA$, whose entry $A_{ij}$ is $1$ if there is an edge between nodes $i$ and $j$, and $0$ otherwise. 
The graph Laplacian of the graph $\mathcal{G}$ is defined as $\bL = \bD - \bA$, where  $\bD = \operatorname{diag}(\bA \bone)$ is the diagonal matrix of (weighted) node degrees, i.e., $D_{ii}$ is the degree of node $i$.
Given an arbitrary orientation of the edges, an alternative description of the structure of $\mathcal{G}$ is the so-called \emph{incidence} matrix $\bB \in \mathbb{Z}^{n \times m}$, such that $B_{ie} = 1$ if node $i$ is the tail of edge $e$, $B_{ie} = -1$ if $i$ is the head of the edge $e$, and $0$ otherwise. 
Using the operator $\bB$, another expression for the graph Laplacian is $\bL = \bB \bB^\top$. 
Notice that we think of the edge $e = (i,j)$ as oriented from the tail $i$ to the head $j$, although the choice of the orientation is arbitrary and has nothing to do with a directed graph.
For notational simplicity, we focus on unweighted graphs, although the presented ideas can be generalized.
For instance, the entry $A_{ij}$ of the adjacency matrix then simply becomes the weight of the edge from $i$ to $j$ if it exists.

To extend the idea of filtering operations to graphs, equation~\eqref{eq:shift_convolution} provides a natural starting point.
In particular, to define a linear filtering operation for general graphs, we may replace the cyclic shift operator $\bS$ by any linear operator encoding the structure of the graph, e.g., the adjacency matrix $\bA$ or the graph Laplacian~$\bL$.
These matrices are accordingly referred to as graph shift operators in the context of GSP~\cite{segarra2017optimal}.
Since linear filtering in the time-domain is defined in terms of convolution, replacing the cyclic shift $\bS$ in the filter~\eqref{eq:shift_convolution} by a graph shift operator $\bG$ gives rise to a graph convolutional filter:
\begin{equation}
    \bs_\text{out} = \sum_{k=0}^{n-1} c_k \bG^k \bs_\text{in} = \bH_\bG \bs_\text{in}.
\end{equation}
We remark that this definition also implies that any filtering operation can be implemented via localized computations in the graph, e.g., by exchanging node signals in a scheme akin to~\eqref{eq:filter_state_space}.
In particular, if we choose a normalized adjacency matrix as the graph shift operator $\bG$, we can see that this filtering scheme is interpretable as a weighted diffusion process on a graph. 
More generally, the graph filtering operation can always be understood in terms of a suitably defined linear dynamical process on the network. 
This highlights again the close similarities between signal processing and dynamical systems.
There is a noteworthy difference in terms of the goals we typically have in mind in this context, however.
In the study of (linear) dynamics on graphs, e.g., diffusion processes, we are often concerned with a fixed dynamics and aim to understand how the properties of an arbitrary graph may influence the behavior of this dynamics.
In contrast, in the context of signal processing, we typically consider the graph as a fixed entity and our goal is to design a filter (or equivalently a dynamical process) that achieves a desired behavior in terms of filtering.

At this point, two natural questions are: (i) What is the influence of the choice of the graph shift operator? (ii) Is there an advantage of choosing one over another graph shift operator?
Let us concentrate on the first question for now.
Since we are concerned with graph filters that can be expressed as a polynomial (or more generally as a power series) of the graph shift operator, the choice of the shift operator will induce an orthogonal basis in which we represent the graph signal $\bs$. 
More precisely, given the spectral decomposition of the symmetric shift operator $\bG = \bU \bLambda \bU^\top$, we say that the eigenvectors of the graph shift operator (the columns of $\bU$) define a \emph{Graph Fourier Transform} (GFT), i.e., they form a natural basis in which the signal $\bs$ can be expressed~\cite{Sandryhaila2013,Ortega2018,Shuman2013}.
The GFT of a signal is defined as
\begin{align}\label{eq:GFT}
    \tilde{\bs} = \bU^\top \bs,
\end{align} 
and the inverse Fourier Transform operation is 
\begin{align}\label{eq:inverse_GFT}
    \bs = \bU \tilde{\bs}.
\end{align}
Given a weight function $h: \mathbb{R} \rightarrow \mathbb{R}$, any shift-invariant graph filter can thus also be written as 
\begin{align}\label{eq:expression_filter}
    \bH_\bG  =  \sum_{k=1}^N h(\lambda_k) \bu_k \bu_k^\top = \bU h(\bLambda) \bU^\top,
\end{align}
where we used the shorthand notation $h(\bLambda) = \operatorname{diag}(h(\lambda_1), h(\lambda_2), \cdots, h(\lambda_n)$).  
Note that this is exactly equivalent to the case of a time-invariant filter, apart from the choice of a different set of basis functions in which the signal is expressed.
Indeed, $h(\bLambda)$ can again be interpreted as the frequency response of the filter $\bH_\bG$, and the filtering operation $\bs_\text{out} = \bH_\bG \bs_\text{in}$ can be decomposed into three steps: i)~express the signal in the Graph Fourier domain via multiplication by $\bU^\top$; ii)~filter the signal by multiplication with $h(\bLambda)$, and iii)~project the filtered signal back into the graph domain via $\bU$.

Based on this discussion, are there any advantages of choosing a particular graph shift operator over another?
Potentially, yes.
As the spectral decompositions of the various possible choices for a graph shift $\bG$ can vary significantly, our choice can have a significant effect.
However, whether the basis functions induced by the graph shift have favorable characteristics for the task at hand is largely dependent on the application context, and there is no choice that will be generally superior in all contexts.
Nonetheless, we will focus primarily on the (combinatorial) graph Laplacian as our graph shift operator.
This choice is on one hand motivated by the mathematical properties of the Laplacian.
On the other hand, as we will see later, the Laplacian can be generalized in a natural way to simplicial complexes, which allows us to draw parallels to the processing of signals on graphs.

Importantly, the graph Laplacian $\bL$ is positive semidefinite, and thus all its eigenvalues are real and non-negative, which enables us to interpret them as frequencies.
In particular, we can order the GFT basis vectors (eigenvectors) according to these frequencies.
This ordering indeed captures the amount of signal variation along the edges of the graph as we can see by considering the eigenvalues in terms of the Rayleigh quotient $$r(\bs) = \frac{\bs^\top \bL \bs }{\bs^\top \bs} =  \frac{\sum_{ij} A_{ij} (s_i- s_j)^2 }{ 2 \|\bs\|^2}.$$
It follows, that eigenvectors associated with small eigenvalues have small variation along the graph edges (i.e., low frequency) and eigenvectors associated with large eigenvalues show  large variation along edges (i.e., high frequency).
Specifically, eigenvectors with eigenvalue $0$ are constant over connected components, akin to a constant signal in the time domain.

\subsection{Illustrative applications of GSP}
To make our discussion of signal processing on graphs more concrete, we consider three application scenarios.
Later, we will use these applications as guiding examples to illustrate how the ideas of signal processing on graphs can be translated to simplicial complexes.

\subsubsection{Signal smoothing and denoising}
Consider a true signal $\by_0$ of interest supported on the set of nodes $\mathcal{V}$.
In many settings, we only observe a noisy version $\by$ of it, i.e., $\by = \by_0 + \boldsymbol{\epsilon} \in \mathbb{R}^N$, where $\boldsymbol{\epsilon}$ is a vector of zero-mean white Gaussian noise. 
For instance, we may consider that $\by_0$ corresponds to a measurement of a sensor in a network, or an opinion of a person in a social network.
Our goal is now to recover the true signal $\by_0$, a procedure that is called \emph{denoising} or \emph{smoothing} in GSP~\cite{chen2014signal,chen2015signal,onuki2016graph}. 

To make this problem well posed, we assume that the signal is \emph{smooth} with respect to the graph structure, i.e., nodes that are connected should have a similar signal~\cite{dong_2016_laplacian, kalofolias2016learn}. 
This assumption translates into the optimization problem
\begin{align}\label{eq:graph_denoising}
    \min_{\hat{\by}} \{ \norm{\hat{\by} - \by }_2^2 + \alpha \hat{\by}^\top \bL \hat{\by}\},
\end{align}
where $\hat{\by}$ is the estimate of the true signal $\by_0$.
The coefficient $\alpha > 0$ can be interpreted as a regularization parameter that trades off the smoothness promoted by minimizing the quadratic form $\hat{\by}^\top \bL \hat{\by} =  \sum_{ij} A_{ij} (\hat{y}_{i} -  \hat{y}_{j})^2/2$ and the fit to the observed signal in terms of the squared $2$-norm.
The optimal solution for \eqref{eq:graph_denoising} is given by 
\begin{align}\label{eq:graph_opt_sl1}
    \hat{\by} = (\bI + \alpha \bL)^{-1} \by = \bH_1 \by.
\end{align}
Note, in particular, that $\bH_1$ is a linear graph filter.

We can also obtain an estimate of the signal using the iterative smoothing operation
\begin{align}\label{eq:graph_opt_sl2}
    \hat{\by} = (\bI - \mu \bL)^{k} \by= \bH_2 \by,
\end{align}
for a certain fixed number of iterations $k$ and a suitably chosen update parameter $\mu$.
This may be interpreted in terms of $k$ gradient descent steps, i.e., discretized gradient flow dynamics, for the potential function $\hat{\by}^\top \bL \hat{\by}$ defining the regularization cost.

Both the denoising operator $\bH_1$ and the smoothing operator $\bH_2$ defined in \eqref{eq:graph_opt_sl1} and \eqref{eq:graph_opt_sl2} are instances of \emph{low-pass filters}, i.e., the frequency responses $h(\boldsymbol{\lambda}) = \operatorname{diag}(\bU^\top \bH \bU)$ are vectors of non-increasing (decreasing) values~\cite{segarra2017optimal}.
Since eigenvectors with small eigenvalues show smaller variation along the graph edges, the low-pass filtering operation guarantees that variations over neighboring nodes are smoothed out.
This is precisely in line with the intuition underpinning the optimization problem~\eqref{eq:graph_denoising}.

\subsubsection{Graph signal interpolation}
Given signal values, called \emph{labels}, for a subset of the nodes $\mathcal{V}^L \subset \mathcal{V}$ of a graph, another common task in GSP is to interpolate the signal on unlabeled nodes $\mathcal{V}^U = \mathcal V \;\backslash\; \mathcal V^L$ in the graph~\cite{narang2013signal, segarra2016reconstruction}.
Similar to the signal denoising problem, we assume that connected nodes have similar labels, which translates into the following optimization problem:
\begin{align}\label{eq:graph_SSL}
    &\min_{\hat{\by}} \norm{\bB^\top \hat{\by}}^2_2, \text{ subject to } \hat{y}_i = y_i \text{ for all } v_i \in \mathcal{V}^L.
\end{align}
Thus, we again aim to minimize the sum-of-squares label difference between connected nodes, but now under the constraint that all observed node labels $y_i$  should be kept in the optimal solution.
Importantly, we assume here that these measurements are fully accurate.
Further, notice that the objective function in \eqref{eq:graph_SSL} can again be written in terms of the quadratic form of the graph Laplacian $\norm{\bB^\top \hat{\by}}^2_2=\hat{\by}^\top \bL \hat{\by}$, highlighting again the inherent low-pass modeling assumption.

\subsubsection{Graph neural networks}

A recent extension to the domain of GSP is the introduction of graph neural networks~\cite{bronstein2017geometric} (GNNs) into the processing pipeline.
In contrast to standard graph filters, GNN architectures include nonlinearities and learnable weights.
GNNs can be built for a range of different tasks including node classification \cite{kipf2016semi, defferrard2016convolutional}, graph classification \cite{gama2018convolutional}, and link prediction \cite{zhang2018link}.
In intuitive terms, one may think of a graph neural network as a procedure to \emph{automatically} find a nonlinear filter (or dynamical process) that fits the desired behaviour given by a training set of graph signals and the desired outputs of such a filter on those signals.

One popular architecture is the graph convolutional network (GCN)~\cite{kipf2016semi}, a generalization of the well-known convolutional neural network architecture for Euclidean data such as time series or images.
A GCN can be understood as a form of iterative smoothing \eqref{eq:graph_opt_sl2} with interleaved element-wise nonlinearities ---usually called activation functions in this context--- and learnable weights that perform linear transformations in the feature space
\begin{align}\label{eq:gnn-layer}
    \bY_{k+1} & = \sigma \left( \bH \bY_k \bW_{k+1} \right)
\end{align}
We run the network for $K$ iterations and define $\bY_K$ as the output of the graph convolution.
The input features $F_0$ are collected in the columns of $\bY_0 \in \mathbb{R}^{N \times |F_0|}$.
Here, $\bH$ is a shift-invariant graph filter based on the graph's adjacency or Laplacian matrix, which is adapted to the task at hand via a set $\left\{ \bW_k \in \mathbb{R}^{|F_{k-1}| \times |F_k|} \right\}_{k=1}^K$ of learnable weight matrices.

Note that we can interpret the GNN in terms of a dynamical system, if we treat each layer in the GNN as corresponding to one time-step.
For simplicity, let us first consider the case where $\sigma(\cdot)$ is the identity mapping.
Then, \eqref{eq:gnn-layer} can be expressed as a linear graph filter that is applied to each feature individually and whose outputs are linearly combined at each node using the matrices $\left\{ \bW_k \right\}$.
That is,
\begin{align}
    \bY_{K} = \bH^K \bY_0 \bW_1 \cdots \bW_K,
\end{align}
where $\bH^K$ is itself a shift-invariant graph filter.
From here, the iterative smoothing~\eqref{eq:graph_opt_sl2} can easily be recovered by restricting $\bY_k$ to only one feature and setting $\bH := \bI - \mu \bL$.
The key benefit of GCNs, however, lies in the interleaved nonlinearities and the linear combination of different features, which enables the network to learn more sophisticated relationships between nodes based on their neighborhoods and node features.

Recurrent graph neural networks take this idea of a nonlinear aggregation of information across neighborhoods to the extreme.
In the basic case, their state evolution can be described as $$\bY_{k+1} = \sigma \left( \bH \bY_k \bW \right)$$ and iterations continue until a stable equilibrium for all node states is reached \cite{wu2020comprehensive}, or some other predefined stopping criterion is fulfilled.
Notice that, unlike in~\eqref{eq:gnn-layer}, the same (learnable) weights are shared across all layers, whose number does not have to be fixed a priori.

\section{Signal processing on simplicial complexes}\label{section:SP_SC}
In this section, we extend our analysis of graph signal processing to the case where signals are not only supported on nodes, but also on higher-order structures such as edges, triangles, and so on. 
One way to analyze such type of data is to model structures via \emph{simplicial complexes} (SC). 
We first briefly review the formalism of simplicial complexes. 
We then discuss how the Hodge Laplacian provides an extension of the graph Laplacian for higher-order networks that can serve as a shift operation for SCs, and enables us to extend denoising and interpolating methods for signals on higher-order networks. 

\subsection{Brief recap of simplicial complexes}\label{section:SP_SC:recap}

Given a finite set of vertices $\mathcal{V}$, a $k$-simplex (or simplex of order $k$) $\mathcal{S}^k$ is a subset of $\mathcal{V}$ with cardinality $k+1$. A simplicial complex $\mathcal{X}$ is a set of simplices such that for any $k$-simplex $\mathcal{S}^k$ in $\mathcal{X}$, any subset of  $\mathcal{S}^k$ must also be in $\mathcal{X}$.

\begin{figure}[tb]
    \centering
    \includegraphics[width=\textwidth ]{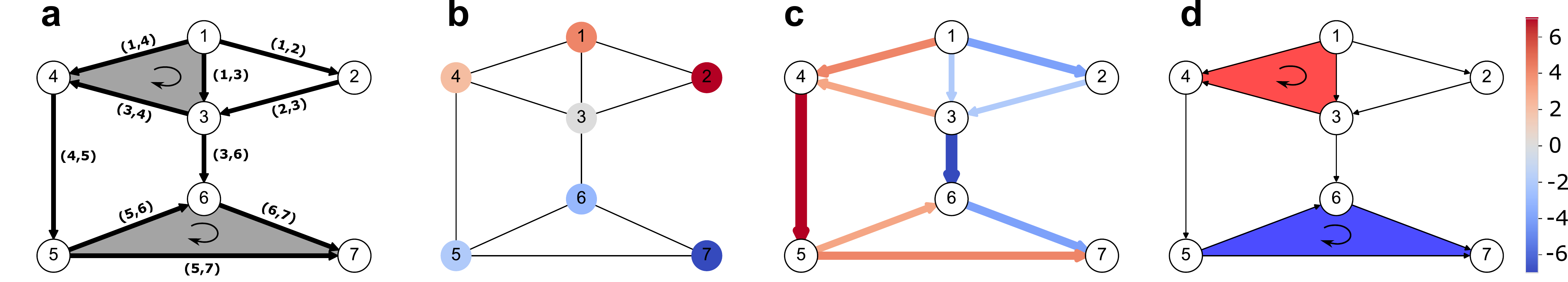}
    \caption{\textbf{Signals on simplicial complexes of different order.} \textbf{a}: Structure of the simplicial complexes used as a running example in the text. Arrows represent the chosen reference orientation. Shaded areas correspond to the $2$-simplices $\{1,3,4\}$ and $\{5,6,7\}$.
    \textbf{b}: Signal on $0$-simplices (nodes).
    \textbf{c}: Signal on $1$-simplices (edges). 
    \textbf{d}: Signal on $2$-simplices (triangles).
    Reproduced from \cite{SPTutorial}.
}
    \label{fig:SC_example}
\end{figure}

\begin{example}
    Consider the simplicial complex given in Figure~\ref{fig:SC_example}a.
    Simplices of order 0 and 1 can be understood as nodes and as edges, respectively.
    Simplices of order 2 correspond to filled triangular faces. 
\end{example}

We can define a relation between $k$-simplices and simplices of order $(k + 1)$ as follows:
We call a $k$-simplex $\mathcal{S}^k$ a \emph{face} of $\mathcal{S}^{k+1}$ if $\mathcal{S}^k$ is a subset of $\mathcal{S}^{k+1}$.
Likewise, $\mathcal{S}^{k+1}$ is a \emph{co-face} of $\mathcal{S}^k$ if $\mathcal{S}^{k+1}$ has exactly one additional element than $\mathcal{S}^k$.

\begin{example}
    Consider again Figure~\ref{fig:SC_example}a.
    The edges $\{1,3\}$, $\{1,4\}$ and $\{3,4\}$ are all faces of the 2-simplex $\{1,23,4\}$.
    The 2-simplex $\{5,6,7\}$ is a co-face of each of the edges $\{5, 6\}$, $\{5, 7\}$ and of $\{6, 7\}$.
\end{example}

To enable algebraic computations with simplicial complexes, we fix an arbitrary ordering of the nodes in the graph.
This ordering induces an orientation for each simplex via the increasing order of the vertex labels. 
Note that this orientation is distinct from the notion of a direction found in a directed graph.
Having defined an orientation for each simplex, we can now keep track of the relationships between simplices of different orders by using linear maps called \emph{boundary operators}.
In our context, these boundary operators are nothing but matrices $\bB_k$, whose rows are indexed by $(k-1)$-simplices and whose columns are indexed by $k$-simplices.
In particular, the $ij$th entry of $\bB_k$ is $+1$ (or $-1$) if the $i$th $(k-1)$-simplex is included in the $j$th $k$-simplex and their orientation is aligned (or anti-aligned), and $0$ otherwise.

\begin{example}
    In Figure~\ref{fig:SC_example}a, we fixed an arbitrary orientation by numbering the nodes from 1 to 7.
    The orientation of 2-simplices is chosen such that nodes appear in increasing order.
    For this simplicial complex, the boundary operators can be represented as follows:
    {\footnotesize \let\quad\thinspace
        \begin{align*}
        \mathbf{B}_1 = \bordermatrix{
    & (1,2)&(1,3)&(1,4)&(2,3)&(3,4)&(3,6)&(4,5)&(5,6)&(5,7)&(6,7) \cr
            1&-1&-1&-1&0&0&0&0&0&0&0\cr
            2&1&0&0&-1&0&0&0&0&0&0\cr
            3&0&1&0&1&-1&-1&0&0&0&0\cr
            4&0&0&1&0&1&0&-1&0&0&0\cr
            5&0&0&0&0&0&0&1&-1&-1&0\cr
            6&0&0&0&0&0&1&0&1&0&-1\cr
        7&0&0&0&0&0&0&0&0&1&1}
    \end{align*}
    \begin{align*}
        \mathbf{B}_2 = \bordermatrix{ & (1,3,4)&(5,6,7)\cr
                    (1,2)&0&0\cr
                    (1,3)&1&0\cr
                    (1,4)&-1&0\cr
                    (2,3)&0&0\cr
                    (3,4)&1&0\cr
                    (3,6)&0&0\cr
                    (4,5)&0&0\cr
                    (5,6)&0&1\cr
                    (5,7)&0&-1\cr
                    (6,7)&0&1}
            \end{align*}}
Note that $\bB_1$ is nothing more than the incidence matrix of the graph and $\bB_2$ is the edge-to-triangle incidence matrix.
\end{example}

\subsection{The Hodge Laplacian as a shift operator for simplicial complexes}\label{subsection:Hodge_Laplacian}
Given signals supported on the $k$-simplices of an SC, we need to define an appropriate shift operator in order to translate the results from the GSP setting to SCs.
To this end, using the boundary operators $\bB_k$ described above, we extend the definition of the graph Laplacian for simplicial complexes. 
The \emph{$k$th combinatorial Hodge Laplacian} is given by~\cite{Eckmann1944,Lim2015}: 
\begin{equation}\label{eq:Hodge_Laplacian}
        \bL_k = \bB_k^\top \bB_k + \bB_{k+1} \bB^\top_{k+1}.
\end{equation}
Note in particular that the graph Laplacian corresponds to $\bL_0 = \bB_1 \bB_1^\top$, with $\bB_0 := 0$. 
Similar to the graph Laplacian, weighted versions of the Hodge Laplacian can be defined as well, but we stick to the unweighted versions here for simplicity.

While we have discussed general SCs so far, to make our discussion more concrete we will focus on signals supported on $1$-simplices, which may be interpreted as edge flows.
This choice is motivated from a practical point of view, as in many application scenarios we are confronted with flows supported on the edges of a network, e.g., in transportation and supply networks, or networks defined via information flows or human mobility.
We argue that in those contexts the $\bL_1$-Hodge Laplacian is a natural choice for a shift operator for signals supported on edges ($1$-simplices).
{The ensuing discussions are still applicable to signals supported on higher-order components of simplicial complexes, which may come up in domains such as electromagnetics~\cite{Deschamps1981}.
Indeed, there are close connections between signals on simplicial complexes and differential forms on manifolds.}

Specifically, similar to the graph Laplacian, the Hodge Laplacian is positive semi-definite, which ensures that we can interpret its eigenvalues in terms of non-negative frequencies.
Moreover, these frequencies are again aligned with a notion of signal-smoothness displayed by the eigenvectors of the Hodge Laplacian.
This notion of smoothness can be understood by means of the so-called $\emph{Hodge decomposition}$~\cite{Lim2015,Grady2010,Schaub2018}, which states that the space of $k$-simplex signals can be decomposed into three orthogonal subspaces
\begin{align}\label{eq:Hodge_decomposition}
    \mathbb{R}^{n_k} =  \im(\bB_{k+1})\oplus \im(\bB_k^\top) \oplus \ker(\bL_k),
\end{align}
where $\im(\cdot)$ and $\ker(\cdot)$ are shorthand for the \emph{image} and \emph{kernel} spaces of the respective matrices, $\oplus$ represents the union of orthogonal subspaces, and $n_k$ is the cardinality of the space of signals on $k$-simplices (i.e., $n_0=n$ for the node signals, and $n_1= |\mathcal{E}|$ for edge signals).
Here we have (i) made use of the fact that a signal on a finite dimensional set of $n_k$ simplices is isomorphic to $\mathbb{R}^{n_k}$; and (ii) implicitly assumed that we are only interested in real-valued signals and thus a Hodge decomposition for a real valued vector space (see~\cite{Lim2015} for a more detailed discussion).

\begin{figure}[tb]
    \centering
    \includegraphics[width = \textwidth]{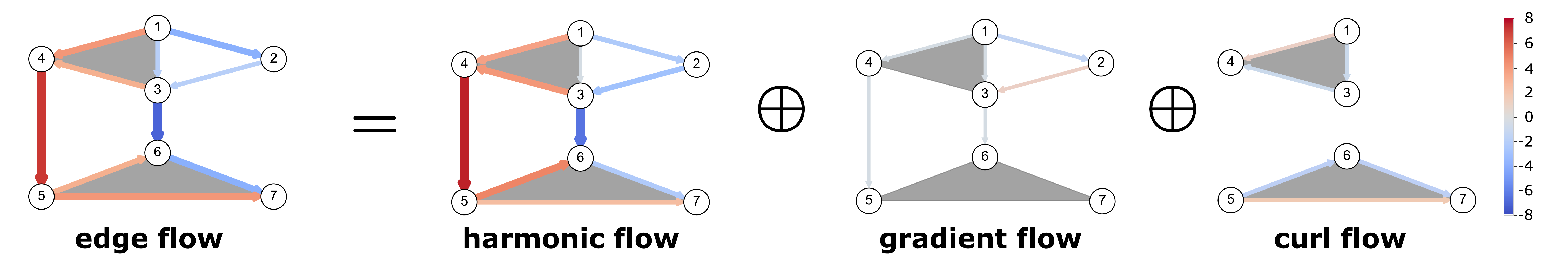}
    \caption{\textbf{Hodge decomposition of the edge flow in the example from Figure~\ref{fig:SC_example}}. Any edge flow (left) can be decomposed into a harmonic flow, a gradient flow, and a curl flow. Reproduced from \cite{SPTutorial}.}
    \label{fig:my_label}
\end{figure}

For the edge-space Hodge Laplacian $\bL_1$ of an SC, this Hodge decomposition is the discrete analogue of the well-known vector calculus result that any vector field can be decomposed into gradient, curl, and harmonic components (see Figure~\ref{fig:my_label} for an illustration.).
First, the space $\im(\bB_1^\top) = \{\mathbf{f} = \bB_1^\top \mathbf{v}, \text{ for some } \mathbf{v} \in \mathbb{R}^{n}\}$ can be considered as the space of gradient flows (or potential flows).
Specifically, we can create any such gradient flow by (i) assigning a scalar potential to all of the $n$ nodes, and (ii) inducing a flow along the edges according to the difference of the potentials on the endpoints.
Clearly, any such flow cannot have a positive net sum along any closed path in the graph and the space orthogonal to $\im(\bB_1^\top)$ is thus called the cycle space.
As indicated by~\eqref{eq:Hodge_decomposition}, the cycle space is spanned by two types of cyclic flows.
The space $\im(\bB_{2})$ consists of curl flows that can be composed of linear combinations of local circulations along any $2$-simplex (triangular face).
Specifically, we may assign a scalar potential to each {oriented} $2$-simplex, and consider the induced flows $\mathbf{f} = \bB_2 \mathbf{t}$, where $\mathbf{t}$ is the vector of $2$-simplex potentials.
Finally $\ker(\bL_1)$ is the harmonic space, whose elements correspond to global circulations that cannot be represented as  linear combinations of curl flows. 

Importantly, these three subspaces are spanned by certain subsets of eigenvectors of $\bL_1$ as described by the following result, which can be verified by direct computation \cite{Barbarossa2020,Schaub2018}.
\begin{theorem}
    Let $\bL_1 = \bB_1^\top \bB_1 + \bB_2 \bB_2^\top$ be the Hodge 1-Laplacian of a simplicial complex.
    The eigenvectors with nonzero eigenvalues of $\bL_1$ consist of two groups that span the gradient space and the curl space, respectively:
    \begin{itemize}
        \item Consider an eigenvector $\bv_i$ of the graph Laplacian $\bL_0$ with  nonzero eigenvalue $\lambda_i$. 
            Then, $\bu_\text{grad}^{(i)} = \bB_1^\top\bv_i$ is an eigenvector of $\bL_1$ with the same eigenvalue $\lambda_i$. 
            Moreover, $\bU_\text{grad} = [\bu_\text{grad}^{(1)},\bu_\text{grad}^{(2)},\ldots]$ spans the space of all gradient flows.
        \item Consider an eigenvector $\mathbf{t}_i$ of the matrix $\mathbf{T} = \bB_2^\top\bB_2$ with nonzero eigenvalue $\theta_i$. 
            Then, $\bu_\text{curl}^{(i)} = \bB_2\mathbf{t}_i$ is an eigenvector of $\bL_1$ with the same eigenvalue $\theta_i$. 
            Moreover $\bU_\text{curl} = [\bu_\text{curl}^{(1)},\bu_\text{curl}^{(2)},\ldots]$ spans the space of all curl flows.
    \end{itemize}
\end{theorem}
The above result shows that, unlike in the case of node signals, edge-flow signals can have a high frequency contribution due to two different types of (orthogonal) basis components being present in the signal: a high frequency may arise both due to a curl component as well as a strong gradient component present in the edge-flow.
This has certain consequences for the filtering of edge signals that we will discuss in more detail in the following section.

\subsection{Illustrative applications} \label{section:examples_SC}
In the following subsections, we revisit the three application scenarios outlined in the context of graphs, but this time focusing on edge flows supported on general SCs.
As we will see, in this context the Hodge Laplacian becomes a natural substitute for the graph Laplacian.
While most of the mathematical formulations can be carried out in essentially the same way when using this substitution, it is important to consider how the interpretation of smoothing and denoising changes when using the Hodge Laplacian in the edge space as a shift operator.

\subsubsection{Flow smoothing and denoising}
Let us reconsider the problem of smoothing and denoising for oriented edge-signals $\bef^0 \in \mathbb{R}^E$ (flows) supported on a simplicial complex $\mathcal{X}$. 
Let us assume again that we cannot observe these flows directly, but we get to see a noisy signal $\bef = \bef^0 + \boldsymbol{\epsilon}$, where $\boldsymbol{\epsilon}$ is a zero-mean white Gaussian noise vector.
As in the case of node signals, our objective is to recover the true underlying signal $\bef^0$.
By analogy with the estimation problem on graphs, we consider solving the following optimization program
\begin{align}\label{eq:edge_denoising}
    \min_{\hat{\bef}} \left\{ \norm{\bfh - \bef }_2^2 + \alpha \bfh^\top \bQ \bfh \right\},
\end{align}
with optimal solution $\bfh = \bH_Q\bef := (\bI + \alpha\bQ)^{-1}\bef$. 
Like before, the quadratic form $\bfh^\top \bQ \bfh$ acts again as regularizer.
Since the filter $\bH_Q$ will inherit the eigenvectors of the matrix $\bQ$, the eigenvectors will form a canonical basis for the filtered signal.
A natural choice for a regularizer is thus an appropriate (simplicial) shift operator. 

Here we discuss three possible choices for the regularizer (shift operator) $\bQ$: (i) the graph Laplacian $\bL_\text{LG}$ of the line-graph~\cite{Schaub2018a} of the underlying graph skeleton of the complex $\mathcal X$, i.e., the line-graph of the graph induced by the $0$-simplices (nodes) and $1$-simplices (edges) of $\mathcal X$; 
(ii) the edge Laplacian $\bL_e= \bB_1^\top \bB_1$,  i.e., a form of the Hodge Laplacian that ignores all $2$-simplices in the complex $\mathcal X$ such that $\bB_2=0$; 
(iii) the Hodge Laplacian $\bL_1 = \bB_1^\top \bB_1 + \bB_2 \bB_2^\top$ that takes into account all the triangles of $\mathcal X$ as well.
To gain some intuition, let us illustrate the effects of these choices by means of an example.

\begin{example}\label{ex:edge_smoothing}
    Figure~\ref{fig:flow_smoothing}a displays a conservative cyclic flow on the edges of an SC, i.e., all of the flow entering a node exits the node again.
This flow is then distorted by a Gaussian noise vector $\boldsymbol{\epsilon}$ in Figure~\ref{fig:flow_smoothing}b. The estimation error produced by the filter based on the line-graph (Figure~\ref{fig:flow_smoothing}c) is comparatively worse  than the estimation performance of the edge Laplacian (Figure~\ref{fig:flow_smoothing}d) and the Hodge Laplacian (Figure~\ref{fig:flow_smoothing}e) filters:
Specifically the 2-norm of the error is $36.54$ (line graph) vs. $1.95$ (edge Laplacian), and $1.02$ (Hodge Laplacian) respectively.
\begin{figure}
    \centering
    \includegraphics[width = \textwidth]{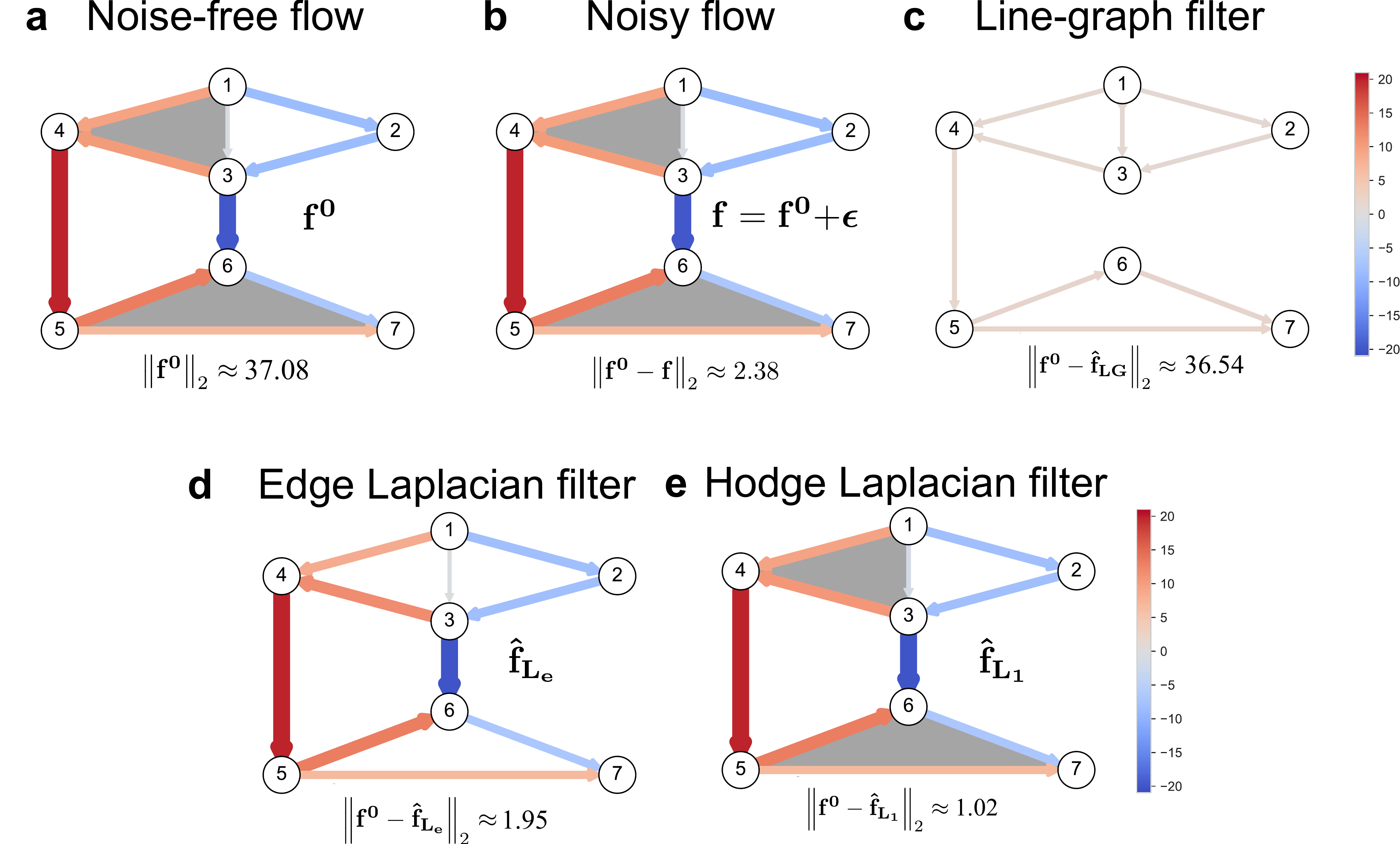}
    \caption{\textbf{Flow smoothing on an SC.} \textbf{a} An SC with a pre-defined and oriented flow $\bef^0$. \textbf{b} The observed flow is a noisy version of the flow $\bef^0$, i.e., $\bef^0$ is distorted by a Gaussian white noise vector~$\boldsymbol{\epsilon}$. \textbf{c} We denoise the flow by applying a Laplacian filter based  on the line-graph. This filter performs worse compared to the edge space filters in \textbf{d} and \textbf{e} that account for flow conservation. \textbf{d} Denoised flow obtained after applying the filter based on the edge Laplacian. \textbf{e} Denoised flow obtained after applying the filter based on the Hodge Laplacian. The estimation error is lower than in the edge Laplacian case as the filter accounts for filled faces in the graph. Reproduced from \cite{SPTutorial}.}
    \label{fig:flow_smoothing}
\end{figure}
\end{example}

To explain the results obtained from the individual filters in the above example, it is essential to realize that the eigenvectors of the regularizer $\bQ$ associated with small eigenvalues will incur a small regularization penalty.
In the case of the line-graph Laplacian, these eigenvectors are determined by the edge connectivity: the low-frequency eigenvectors correspond to signals in which adjacent edges in the simplicial complex have a small difference.
This is equivalent to the notion that low-frequency modes in the node space do not vary a lot on tightly connected nodes on a graph.
However, for flow signals this notion of smoothness induced by the line-graph as shift operator is often inappropriate.
Specifically, many real-world flow signals are approximately conservative, in that most of the flow signal entering a node exits the node again, but the relative allocation of the flow to the edges does not have to be similar.
Accordingly, as can be seen from Figure~\ref{fig:flow_smoothing}c, the line graph filtering operation leads to an increased error compared to the noisy input signal~\cite{Schaub2018a}. 
Another reason for this behavior is that the line-graph Laplacian does not reflect the arbitrary orientation of the edges, so that the performance is not invariant to the chosen sign of the flow.

Unlike the line-graph Laplacian, the Edge Laplacian captures a notion of flow conservation, and its zero frequency eigenvectors correspond to cyclic flows~\cite{Schaub2018a}.
To see this, it is insightful to inspect the quadratic regularizer ${\bL_e = \bB_1^\top\bB_1}$.
Note that this quadratic form can be written as ${\bef^\top \bL_e\bef = \|\bB_1 \bef\|_2^2}$.
This is precisely the (summed) squared divergence of the flow signal $\bef$, as each entry $(\bB_1 \bef)_i$ corresponds to the difference of the inflow and outflow at node~$i$.
As a consequence, all cyclic flows will induce zero cost for the regularizer $\bef^\top \bL_e\bef$.
Stated differently, any flow that is \emph{not} divergence free, i.e., not cyclic, will be penalized by the quadratic form.
Since by the fundamental theorem of linear algebra $\ker(\bB_1) \perp \im (\bB_1^\top)$, any such non-cyclic flow can be written as a gradient flow $\bef_\text{grad} =\bB_1^\top \mathbf{v}$ for some vector $\mathbf{v}$ of scalar node potentials --- in line with the Hodge decomposition discussed in~\eqref{eq:Hodge_decomposition}.

In contrast to the Edge Laplacian, the full Hodge Laplacian $\bL_1$ includes the additional regularization term $\bef^\top \bB_2\bB_2^\top \bef = \|\bB_2^\top\bef\|_2^2$, which may induce a non-zero cost even for certain cyclic flows. 
More precisely, any curl flow $\bef_\text{curl}=\bB_2\mathbf{t}$, for some vector $\mathbf{t}$ will have a non-zero penalty.
This penalty is incurred despite the fact that $\bef_\text{curl}$ is a cyclic flow by construction: since $\bB_1\bef_\text{curl} = \bB_1\bB_2\mathbf{c} = 0$, the vector $\bef_\text{curl}$ is clearly in the cycle space; see also discussion in Section~\ref{subsection:Hodge_Laplacian}.
The additional regularization term $\|\bB_2^\top\bef\|_2^2$ may thus be interpreted as a squared curl flow penalty.

From a signal processing perspective, the $\bL_1$ based filter thus allows for a more refined notion of a smooth signal. 
Unlike in the Edge Laplacian filter, not all cyclic flows can be constructed from frequency (eigenvalue) $0$ basis signals.
Instead a signal can have a high-frequency even if it is cyclic, when it has a high curl component.
Hence, by constructing simplicial complexes with appropriate (triangular) $2$-simplices, we have additional modeling flexibility for shaping the frequency response of an edge-flow filter.
In our example above, this is precisely what leads to an improvement in the filtering performance.
Indeed the displayed ``ground truth'' signal is a harmonic function with respect to the simplicial complex and thus does not contain any curl components.
We remark that the eigenvector basis of $\bL_e$ can always be chosen to be identical to the eigenvectors of $\bL_1$; thus, we may represent any signal in exactly the same way in a basis of $\bL_e$ or $\bL_1$.
Thus the difference in filtering performance is not due to the chosen eigenvector basis, but only due to the eigenvalues:
the frequencies associated with all cyclic vectors will be $0$ for the Edge Laplacian, while there will be cyclic flows with nonzero frequencies for $\bL_1$, in general.
This emphasizes that the construction of faces is an important modeling choice when defining the appropriate notion of a smooth signal.

\subsubsection{Interpolation and semi-supervised learning}
We now focus on the interpolation problem for data supported on the edges of a simplicial complex. 
Let us suppose that we measure signals on a subset of the edges in $\mathcal{X}$, i.e., we are given a set of labeled edges $\mathcal{E}^L \subset \mathcal{E}$, with cardinality $|\mathcal{E}^L| = E_L$. 
The objective is to estimate the labels of unobserved or unmeasured edges in the set $\mathcal{E}^U \equiv \mathcal{E}\backslash \mathcal{E}^L$, whose cardinality we will denote by $|\mathcal{E}^U|=E_U$. 
Following~\cite{Jia2019}, we will again start by considering the problem setup with no $2$-simplices ($\bB_2=0$), before we consider the general case in which $2$-simplices are present.

To derive a well-defined problem for imputing the remaining edge-flows, we need to assume that the true signal has some low-dimensional structure.
Following our above discussions, we will again assume that the true signal has a low-pass characteristic in the sense of the Hodge $1$-Laplacian, i.e., that the edge flows are mostly conserved. 
Let $\bfh$ denote the vector of the true (partly measured) edge-flow.
A convenient loss function to promote flow conservation is then again the sum-of-squares vertex divergence
\begin{align}\label{eq:vertex_divergence}
    \norm{\bB_1 \bfh}^2_2 = \bfh^\top \bB_1^\top \bB_1 \bfh = \bfh^\top \bL_e \bfh.
\end{align}
Accordingly, we formalize the flow interpolation problem 
\begin{align} \label{eq:SC_SSL}
        \min_{\bfh} \norm{\bB_1 \bfh}_2^2 + \alpha^2 \cdot \norm{\bfh}_2^2 \text{ s.t. } \hat{f}_r = f_r, \text{ for all measured edges } r \in \mathcal{E}^L,
\end{align}
Note that, in contrast to the node signal interpolation problem, we have to add an additional regularization term $\|\bfh \|_2^2$ to guarantee the uniqueness of the optimal solution.
If there is any independent cycle in the network for which we have no measurement available, we may otherwise add any flow along such a cycle while not changing the divergence in~\eqref{eq:vertex_divergence}.
To remedy this aspect, we simply add a $2$-norm regularization that promotes small edge-flow magnitudes by default.
While other regularization terms are possible, with this formulation we can rewrite the above problem in a least squares form as described next.

To arrive at a least-squares formulation, we consider a trivial feasible solution $\bfh^0$ for~\eqref{eq:SC_SSL} that satisfies $\hat{f}^0_r = f_r$ if $r \in \mathcal{E}^L$ and $\hat{f}^0_r = 0$ otherwise. 
Let us now define the expansion operator $\bm \Phi$ as the linear map from $\mathbb{R}^{E_U}$ to $\mathbb{R}^{E}$ such that the true flow $\bef$ can be written as $\bef = \bfh^0 + \bm\Phi \bef^U$, where $\bef^{U}\in \mathbb{R}^{E_U}$ is the vector of the unmeasured true edge-flows.
Reducing the number of variables considered in this way, we can convert the constrained optimization problem \eqref{eq:SC_SSL} into the following equivalent unconstrained least-squares estimation problem for the \emph{unmeasured} edges $\bfh^{U}$:
\begin{align}\label{eq:SSL_Le}
    \bfh^{U*} = \argmin_{\bfh^U} \norm{ \left[\begin{array}{c}  \bB_1 \bm \Phi\\ \alpha \bI \end{array} \right] \bfh^U - \left[\begin{array}{c}  -\bB_1\bef^0\\ 0 \end{array} \right] }^2_2.
\end{align}
We illustrate the above procedure by the following example.
\begin{example}
We consider the network structure in Figure~\ref{fig:SC_example}a. The ground truth signal is $\bef = [-2,-2,4,-2,3,-7,7,3,4,-4]^{\top}$. We pick five labeled edges at random (colored in Figure~\ref{SSL_edge_flow}a). The goal is to predict the labels of the unlabeled edges (in grey with a question mark in Figure~\ref{SSL_edge_flow}a). The set of labeled edges is $\mathcal{E}^L = \{(1,3),(1,4),(3,6),(4,5),(5,6)\}$. The set of unlabeled edges is $\mathcal{E}^U = \{(1,2),(2,3),(3,4),(5,7),(6,7)\}$. Solving the optimization program \eqref{eq:SSL_Le}, we obtain the predicted signal $\mathbf{f}^*_{SSL}$ in Figure~\ref{SSL_edge_flow}b.
Numerical values are given in Figure~\ref{SSL_edge_flow}c.
The Pearson correlation coefficient between $\bef$ and $\mathbf{f}^*_{SSL}$ is 0.99. The $2$-norm of the error is 0.064.

\begin{figure}[tb]
    \centering
    \includegraphics[width = \textwidth]{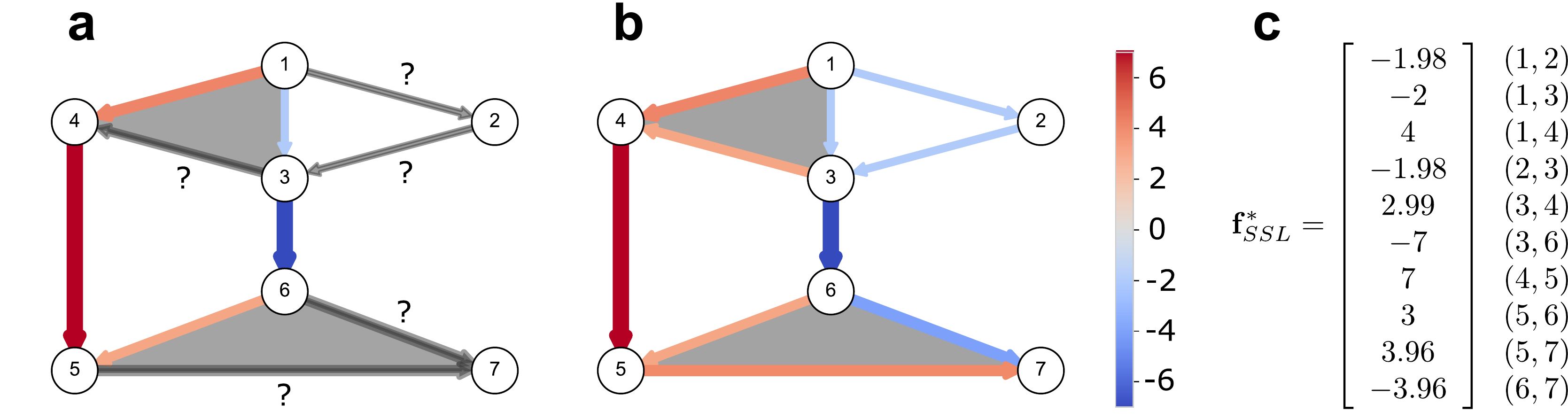}
    \caption{\textbf{Semi-supervised learning for edge flow.} \textbf{a} Synthetic flow. 50\% of the edges are labeled. Labeled edges are colored based on the value of their flow. The remaining edges in grey are inferred from the procedure explained in the text. \textbf{b} Edge flow obtained after applying the semi-supervised algorithm in~\eqref{eq:SSL_Le}. \textbf{c} Numerical value of the inferred signal. Reproduced from \cite{SPTutorial}.}
    \label{SSL_edge_flow}
\end{figure}
\end{example}

Analogous to our discussion above, it may be relevant to include $2$-simplices for the signal interpolation problem.
We interpret such an inclusion of $2$-simplices in two ways.
From the point of view of the cost function, it implies that instead of penalizing primarily gradient flows (which have nonzero divergence), we in addition penalize certain cyclic flows, namely those that have a nonzero curl component.
From a signal processing point of view, it means that we are changing what we consider a smooth (low-pass) signal, by adjusting the frequency representation of certain flows.
Accordingly, one possible formulation of the signal interpolation problem, including information about $2$ simplices is
\begin{align}
\bfh^{\star} &= \argmin_{\bfh} \norm{\bB_1 \bfh}_2^2 + \norm{\bB_2^\top \bfh}_2^2 + \alpha^2 \norm{\bfh}_2^2,
\end{align}
subject to the constraint that the components of $\bfh$ corresponding to measured flows are identical to those measurements.
As in \eqref{eq:SSL_Le}, we can convert this program into the following least-squares problem
\begin{align}
    \bfh^{U \star} &= \argmin_{\bfh^U} \norm{ \left[\begin{array}{c}  \bB_1 \bm\Phi\\ \alpha \bI \\ \bB_2^\top \bm\Phi \end{array} \right] \bfh^U - \left[\begin{array}{c}  -\bB_1 \bef^0 \\ 0 \\ - \bB_2^\top \bef^0\end{array} \right] }^2_2.
\end{align}

\begin{remark}
    Note that the problem of flow interpolation is tightly coupled to the issue of signal reconstruction from sampled measurements.
    Indeed, if we knew that the edge signal to be recovered was exactly bandlimited~\cite{Barbarossa2020}, then we could reconstruct the edge-signal if we had chosen the edges to be sampled appropriately.
    Just like the interpolation problem considered here may be seen as a semi-supervised learning problem for edge labels, finding and choosing such optimal edges to be sampled may be seen as an active learning problem in the context of machine learning.
    While we do not expand further on the choice of edges to be sampled here, we point the interested reader to two heuristic active learning algorithms for edge flows presented in \cite{Jia2019}. 
    We also refer the reader to \cite{Barbarossa2018,Barbarossa2020} for a theory of sampling and reconstruction of bandlimited signals on simplicial complexes, and to \cite{barbarossa2020spmag} for a similar overview that includes an approach for \emph{topology inference} based on signals supported on simplicial complexes.
\end{remark}

\subsubsection{Simplicial neural networks}

With a foundation in linear filtering based on the Hodge Laplacian for $k$-simplex signals, a natural next step is the interleaving of nonlinearities to form convolutional neural networks for data on simplicial complexes.
In particular, convolutional layers analogous to~\eqref{eq:gnn-layer} can be constructed by composing filters, boundary maps, and activation functions.
Letting $\bY_0\in\mathbb{R}^{n_k\times|F_0|}$ gather $k$-simplex signals in its columns, the layers of a simple convolutional neural network, as done by~\cite{ebli2020simplicial}, can be written recursively as
\begin{equation}\label{eq:snn-layer}
    \bY_{k+1} = \sigma \left( \bH \bY_k \bW_{k+1} \right),
\end{equation}
where $\bH$ is some suitable polynomial of the Hodge Laplacian $\bL_k$, and each $\bW_{k+1}\in\mathbb{R}^{|F_k|\times|F_{k+1}|}$ is a (learnable) weight matrix.
Variants of this basic architecture allow weights for the upper ($\bB_k^\top\bB$) and lower ($\bB_{k+1}\bB_{k+1}^\top$) components to be learned independently~\cite{glaze2021principled}, or for signals to be computed over all dimensions of the simplicial complex~\cite{bunch2020simplicial}.

When representing a $k$-simplex signal for $k\geq 1$, the first step is to fix an arbitrary orientation of the simplices, as discussed in~Section~\ref{section:SP_SC:recap}.
In doing so, one fixes the signs of the entries in the incidence matrices $\bB_k,\bB_{k+1}$, as well as the signs of a given $k$-simplex signal.
This is a distinct choice compared to architectures for signals on the nodes of a graph, such as graph neural networks, since nodes do not have a notion of orientation associated to them.
Fortunately, the signs of the incidence matrices and the signs of the signal written as a real vector are compatible, so that linear filters based on the Hodge Laplacian commute with the choice of orientation: that is, such filters are \emph{equivariant} to the chosen orientation of the $k$-simplices.
This is an important property, since the orientation is indeed arbitrary: the choice of orientation for $k$-simplex signals is analogous to choosing a coordinate basis in a vector space.
Although a choice of basis for a vector space makes computation intuitive, the ``true'' object is the vector itself, and not the list of coordinates in that basis.

This introduces a new problem for the design of neural networks, since convolutional layers such as~\eqref{eq:snn-layer} are not necessarily equivariant to the choice of orientation.
This was considered by~\cite{Roddenberry2019,glaze2021principled}, both advocating for the use of \emph{odd, elementwise, and continuous} activation functions in neural networks for signals on simplicial complexes.
Such functions are readily shown to yield architectures equivariant to the arbitrary choice of orientation.

\section{Further relations to dynamical systems on graphs and simplicial complexes}
As we have seen in Section~\ref{section:SP_and_LDS}, the graph Laplacian is intimately connected to linear dynamical systems on networks, such as diffusion processes or consensus processes.

In the context of consensus processes~\cite{olfati2007consensus, zhu2020network}, we consider networks composed of so called ``agents'' (entities, devices, people) that are connected by an edge if the corresponding agents interact. 
At each time step $t$, each agent is in a certain \emph{state}.
For instance, in the context of social networks, the \emph{state} of node $i$ at time $t$ can be understood as the \emph{opinion} of the individual $i$ at time $t$.
Let us denote by $s_i(t)$ the state of node $i$ at time $t$ and store the states of all agents at time $t$ in a vector of states (or \emph{opinions} in the social networks context) $\bs(t) = [s_1(t), s_2(t), \cdots, s_n(t)]^\top$.
The dynamics of $s_i$ over time can be expressed with the averaging law
\begin{align}
    \dot{s}_i(t) = - \sum_{v_i \sim v_j} (s_i(t) - s_j(t)),
\end{align}
which we can rewrite as
\begin{align}\label{eq:consensus_graph}
    \frac{d \bs(t)}{dt} = - \bL \bs(t),
\end{align}
where $\bL$ is the graph Laplacian. 

Given an initial condition $\bs(0) =\bs_0$, the solution of the above linear dynamics is simply $\bs(t) = \exp(-\bL t))\bs_0$, where $\exp(\cdot)$ denotes the matrix exponential.
Note we may interpret this vector equivalently as a filtered graph signal.
Stated differently, for every $t>0$, we may interpret the current state vector of the system as a (low-pass) filtered version of the initial condition~$\bs_0$.

Following \cite{Muhammad06controlusing}, we can also study a generalization of the dynamical system in \eqref{eq:consensus_graph} for higher-order Laplacians. 
Let us consider a discrete time-varying signal $\bom(t)$ of order $k$. 
For instance, in the case $k = 1$, $\bom(t)$ is an edge-flow, analog to $\bef$ in Section~\ref{section:examples_SC}, i.e., the $i$-th component of the vector $\bom(t)$ is the state of the $i$-th edge, assuming that an ordering of the edges was initially defined. 
We consider the dynamical system
\begin{align}\label{eq:LDS_SC}
    \frac{d \bom(t)}{dt} = -\bL_k \bom(t),\qquad \bom(0) = \bom_0,
\end{align}
where $\bL_k$ is the \emph{$k$-th combinatorial Hodge Laplacian} as defined in \eqref{eq:Hodge_Laplacian}. 
We can interpret \eqref{eq:LDS_SC} as the higher-order analog of the discretized heat equation given in \eqref{eq:consensus_graph} (note, however, that this is not a diffusion process).
The equilibrium points of this dynamical system are given by the set \cite{Muhammad06controlusing}
\begin{align}
    \{\bom \;|\; \bL_k \bom = 0\} = \ker(\bL_k).
\end{align}

Since $\bL_k$ is a positive semi-definite matrix and the system~\eqref{eq:LDS_SC} is thus (semi-)stable, the dynamics in \eqref{eq:LDS_SC} can be seen as a means to compute a $k$-th order harmonic signal on a generic simplicial complex, starting from any arbitrary $k$-th  order signal.
If the simplicial complex has non-empty homology (many \emph{holes}), then the system will converge to an element in the basis of $\ker(\bL_k)$ depending on the initial condition.
In other words, the above dynamical system acts as a perfect low-pass filter which projects the initial condition (asympotically) into the space of harmonic signals.

In particular, if the simplicial complex has exactly one hole, then for any initial condition, the system will converges to a vector that spans the harmonic subspace. 
The dynamical system in \eqref{eq:LDS_SC} thus offers a decentralized method to compute the \emph{homology classes} in the simplicial complex~\cite{Muhammad2007}. 
This method finds a useful application in real sensor networks \cite{Muhammad2007,Tahbaz2010}, in particular for detecting coverage holes in a decentralized manner and without location information \cite{Tahbaz2010}.

Generalizations of \eqref{eq:LDS_SC} have been discussed and analyzed in \cite{Torres2020,deVille2020}. 
In particular, \cite{deVille2020} analyzes a nonlinear version of \eqref{eq:LDS_SC}. 
For an unweighted SC, this model takes the form:
\begin{align}\label{eq:LDS_SC_nonlinear}
    \frac{d \bom(t)}{dt} = -\left( \bB_{k+1} f(\bB_{k+1}^\top \bom(t)) + \bB_k^\top f(\bB_k \bom(t)) \right),
\end{align}
where $f : \mathbb{R}^n \rightarrow \mathbb{R}^n$ is a function that acts componentwise.
It is shown in \cite{deVille2020} that \eqref{eq:LDS_SC_nonlinear} under certain conditions be formulated as the gradient flow for an energy functional defined by the simplicial complex.
Based on these results, the stability of certain steady states in the nonlinear case can be deduced~\cite{deVille2020}.
Note again that, apart from a lack of adjustable weights, there is again a remarkable similarity between the nonlinear equation~\eqref{eq:LDS_SC_nonlinear} and the nonlinear signal processing approach of the neural network formulation for SCs in~\eqref{eq:snn-layer}. 
In particular, both systems are based on alternating applications of a linear transformation and a point-wise nonlinearity.
Indeed this close similarity of particular neural network architectures and ordinary differential equations has been the focal point for the development of the so-called neural ODE formulations~\cite{NeuralODE}.

\section{Discussion}
Simplicial complexes have emerged as a key modeling framework for abstracting complex systems with higher-order interactions~\cite{Battiston2020,Torres2020,bick2021higher}.
The majority of previous works have focused on studying the structural properties of such complexes, in particular in the context of topological data analysis~\cite{wasserman2018topological}.
More recently, the study of dynamical processes acting on top of such complexes has gained attention.
Here we have discussed signal processing for data supported on simplicial complex as a closely aligned, but different perspective to both of these viewpoints.
Rather than trying to understand a dynamical behavior on a (arbitrary but fixed) simplicial complex, in the context of signal processing we aim to obtain a desired filtering output, based on a given input signal --- which may be interpreted as aiming to \emph{design} a particular dynamics that achieves a desired target specification as close as possible.
We have centered our discussion on the Hodge Laplacian~\cite{Lim2015,Eckmann1944} as a key operator whose spectral decomposition provides a unitary basis for signals supported on simplicial complexes, which is tightly coupled to the structural properties of the underlying SC due to the Hodge decomposition.
Specifically, focusing on edge-flows, we discussed how the Hodge decomposition can be interpreted as a discrete analog of the well-known decomposition of a continuous vector field into gradient, curl, and harmonic components.

Our discussion opens up a number of avenues for future research.
For instance, one pertinent question concerns the ``optimal construction" of simplicial complexes from data and how this affects signal processing supported on SCs.
This problem is only magnified when considering weighted SCs and corresponding weighted Hodge Laplacians, to which most of the theory discussed here can be readily applied as well~\cite{Schaub2018,Lim2015}.
More generally, most of the developed theory can be readily extended to cell complexes, in which the atomic building blocks are not simplices but cells containing any number of nodes.
How to choose an appropriate cell complex representation for a given system is a completely open topic that remains to be explored in more detail.

\subsection*{Acknowledgements}
This work was partially supported by USA NSF under award CCF-2008555. MTS and FF acknowledge partial funding from the Ministry of Culture and Science (MKW) of the German State of North Rhine-Westphalia (NRW R\"uckkehrprogramm).
We thank Lucille Calmon for carefully checking and providing feedback on the manuscript.

\bibliographystyle{spmpsci}
\bibliography{bibliography_signal.bib}
\end{document}